\documentclass[12pt]{iopart}

\usepackage{graphicx}
\usepackage{dcolumn}
\usepackage{bm}
\usepackage{comment}
\usepackage{gensymb}

\begin{document}



\graphicspath{{Figs2/}}




\title[Harmonic analysis of atomic magnetization to calibrate a 3D time-dependent field]{Tri-axial time-dependent magnetic field calibrated \textit{in-situ} by harmonic analysis of adiabatically evolving atomic spins}

\author{Giuseppe Bevilacqua}
\address{DSFTA Università di Siena - Via Roma 56 - 53100 Siena, Italy}

\author{Valerio Biancalana}
\ead{valerio.biancalana@unisi.it}

\address{DSFTA Università di Siena - Via Roma 56 - 53100 Siena, Italy}

\author{Yordanka Dancheva}
\address{DSFTA Università di Siena - Via Roma 56 - 53100 Siena, Italy}
\address{  Aerospazio Tecnologie S.r.l. - 53040 Rapolano Terme (SI), Italy }%

\author{Alessandro Fregosi}
%
\address{ CNR - Istituto Nazionale di Ottica - Via G. Moruzzi 1 - 56124 Pisa (Italy) }%

\date{\today}



\begin{abstract}
We introduce a methodology to calibrate \textit{in situ} a set of coils generating bi- or tri-axial magnetic fields, at frequencies where a calibration performed under static conditions would be inaccurate. 
The methodology uses harmonic analysis of one component of the magnetization of an atomic sample whose spins adiabatically follow an \textit{ad hoc} applied time-dependent field. 
The procedure enables the identification of phases and amplitudes of the coil currents required to produce a dynamic magnetic field with the assigned polarization. This determines coil constants that can be subsequently used to produce arbitrary three-dimensional time-dependent fields.
\end{abstract}

\maketitle

\section{Introduction}
\label{sec:indroduction}
The application of precisely assigned magnetic fields with tailored spatial and temporal distribution is at the core of many precision experiments in modern physics particularly in atomic laser spectroscopy
and quantum optics \cite{wang_arx_24, fallon_pra_20}. Generating well-controlled triaxial, time-dependent magnetic fields is of interest also for other application areas \cite{tajuelo_prappl_23}. 
Solenoids, Helmholtz pairs, and other specifically designed coil arrangements \cite{holmes_sr_19} are commonly used to achieve the desired field structure. Numerically controlled power supplies enable precise adjustments of the magnetic field components and the generation of custom-designed time-dependent magnetic fields.

In instances requiring a two- or three-dimensional time-dependent field, setting its Cartesian and Fourier components with accurate amplitude and relative phase is of paramount importance \cite{gerginov_prappl_19,  bevilacqua_tundress_prl_20, wang_arx_24}. 

The  characterization of magnetic field generators can follow both \textit{a priori} and \textit{a posteriori} approaches. The former involves designing the current distribution (coil shape and current waveform) aiming to produce the appropriate fields based on Maxwell's laws. The \textit{a posteriori} approach involves measuring the actual field and adjusting it using additional coils or varying the driving current(s). This can be essential for accuracy because the magnetic field generators are often surrounded by conducting materials or magnetic shields whose effects cannot be precisely considered in simulations or calculations.
Several kinds of sensors can be used to measure the actual field, each of them coming with its inherent degree of precision, accuracy, spatial and time resolution, robustness etc. The use of Hall-effect or fluxgate sensors is often a favorite choice for their practicality, simple structure, and vector response \cite{gregor_mst_05, weyand_ijaem_05}. However, in some applications, they are not sufficiently precise, do not possess the necessary bandwidth, or are not compatible with existing constraints. 
In this respect, the methodology proposed in this work might help build setups for the calibration of those vectorial sensors.

Laser spectroscopy with its unrivaled precision is an area of research that can require extreme accuracy in magnetic field control. At the same time, it enables the construction of excellent magnetic detectors.
As proposed by Breschi et al.\cite{breschi_apb_14}, an elegant and effective way to control the field with the required accuracy and with detectors easily fitting in atomic spectroscopy setups is based on using the atomic sample itself as a magnetometric sensor. This concept forms the basis of \textit{in-situ} calibration procedures reported by several researchers.

H.Zhang et al. \cite{zhang_epjd_16} presented a method to infer the coil constant (ratio between field and current) with high precision employing a high-performance hyperpolarized He magnetometer.
A similar calibration method was proposed by Yao et al. \cite{yao_ps_19}, who used a hybrid potassium-rubidium magnetometer and analyze the magnetic resonance under varying (swept) magnetic fields.
Chen et al. \cite{chen_aipadv_17} proposed a coil calibration method based on the duration of a $\pi/2$ pulse, precisely determined by the maximization of the initial amplitude of the free-induction-decay signal. The latter was generated by free precessing spins in Xenon gas polarized by collisions with optically pumped Rb vapour. This choice eliminates systematic errors that are possibly introduced by the laser radiation.
Zhao et al. \cite{zhao_jmmm_20} used calibrated coils to obtain vector response from a self-oscillating rubidium optically-pumped magnetometer non-destructively interrogated by far-detuned probe radiation. This approach has similarities with that proposed by G.Zhang et al. \cite{zhang_ieee_19}, which used a forced (Bell and Bloom) magnetometer. In this case, atoms were optically pumped by elliptically polarized radiation, enabling of a vector response from the analysis of orientation and alignment dynamics.
More recently, K.Wang et al. \cite{wang_ieee_23} proposed a coil calibration method applied to a spin-exchange-relaxation-free optically-pumped magnetometer based on the dynamic (transient) response of such system upon application of sudden variation of the magnetic field along the three orthogonal directions. A calibration method based on magnetic induction detection with an accurately aligned pickup coil has been proposed by K.Zhang et al. \cite{zhang_mms_23} for Helmholtz coils operated in a frequency range from few tens to several hundred Hz.

A general feature shared by all the above-mentioned works is to reduce the field measurement to a current (more commonly a voltage) measurement through an opportune set of calibration factors to be calculated or measured. The current-field linear dependence makes it possible to indirectly monitor the field by measuring the voltage across an assigned resistor connected in series with the field-generating coil. Those calibration factors are typically determined under static or quasi-static conditions but can be used also in the case of a time-dependent field, under the hypothesis that the instantaneous field can be inferred from the corresponding instantaneous voltage drop on the series resistor.

The last assumption may be no longer valid in the case of fast-varying fields and currents. Spurious phenomena (e.g. parasitic capacitance of cables and coils) can cause the measured current across the series resistor to be different from that actually flowing in the coil, moreover, unperceived eddy currents may be induced in the coil proximity. Thus the high-frequency terms of the produced field may substantially differ, in amplitude and phase, from those inferred from the voltage drop measured on the monitor resistance \cite{gregor_mst_05, wang_arx_24}.

The motivation that inspired the present work is the interest in studying the dynamics of atomic spins that evolve in (or are driven by) arbitrarily oriented time-dependent magnetic fields.  In this research we have studied \cite{bevilacqua_dualdr_pra_22, fregosi_sr_23} 
and applied \cite{bevilacqua_eiitundr_oe_21} phenomena commonly referred to as \textit{magnetic dressing} \cite{haroche_prl_70}, focusing on cases where the dressing field contains various Fourier and Cartesian components \cite{bevilacqua_dressing_pra_12, bevilacqua_tundress_prl_20}. This research requires a very accurate control of the amplitude and relative phases of the magnetic field components. 

Here, we present an original method to accurately generate a circularly polarized field. This method permits an accurate \textit{a posteriori} determination of the calibration parameters, which can subsequently be used to produce any kind of field polarization, 
over a broad frequency range, beyond the limit at which coil constants determined under static conditions become imprecise. Our method relies on atomic spins that follow adiabatically a time-dependent field and this requirement sets an upper limit to the practically applicable frequencies, which must approximately fall in the audio range. The achieved results will find application in all cases where oscillating field components along diverse axes must be generated with very accurate relative phases and amplitudes.

The calibration methodology is based on generating a magnetic field
ideally made of a time-dependent component that rotates on a plane and a static one oriented perpendicularly to that plane. This field drives the evolution of atomic spins and its imperfections, which may consist of non-perfect perpendicularity of the static field and of elliptical polarization of the time-dependent one, are pointed out by harmonic analysis of the measured signal. 

The developed procedure is correspondingly made of two steps. A preliminary step makes the static field perpendicular to the polarization plane of the time-dependent one. The second one establishes a circular polarization of the latter. 

An optically-pumped atomic sample is interrogated with polarimetric techniques extracting a signal proportional to one component of the macroscopic magnetization. Diverse harmonic components of this signal bring information about the mentioned field imperfections, allowing an inherently \textit{in-situ} calibration procedure to be developed based on Fourier analysis.

The presentation is organized as follows: in Sec.\ref{sec:setup} we describe the experimental setup, in Sec.\ref{sec:principle} we describe (demanding calculus details to an Appendix) the principle of operation of the methodology to control the relative orientation of the static and the time-dependent fields (Sec.\ref{subsec:m1m2}) and to refine the relative phases and amplitudes of the time-dependent field components (Sec.\ref{subsec:epsilonphi}). Demonstrative results are reported and analyzed in the Sec.\ref{sec:results}. A synthesis of the achievements is drawn and shortly discussed in Sec.\ref{sec:conclusion}.

\section{Setup}
\label{sec:setup}
\begin{figure}[ht]
   \centering
      \includegraphics [angle=270, width= 1.1 \columnwidth]{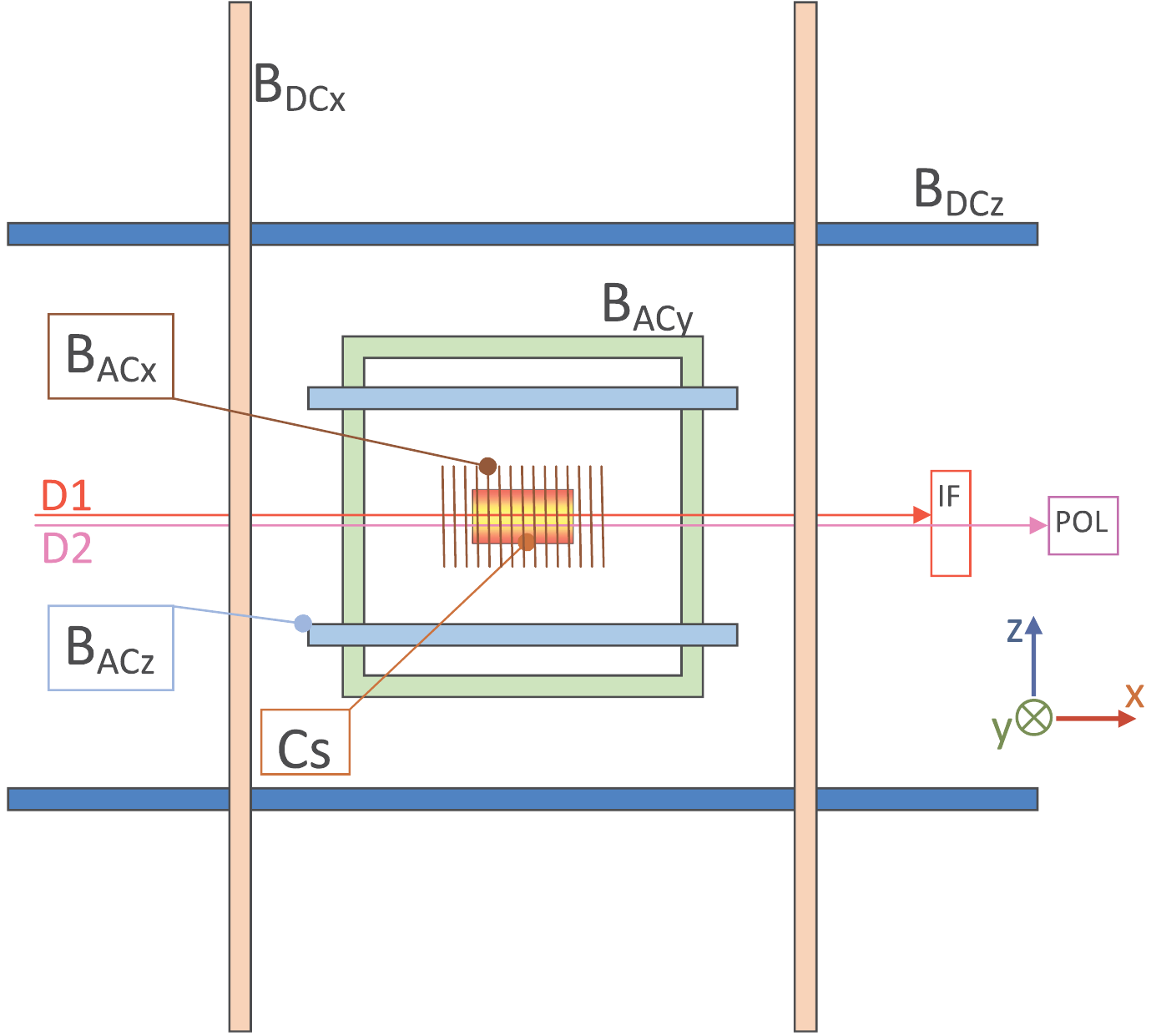}
      \caption {Schematics of the setup. Cs atoms contained in a sealed cell (Cs) are optically pumped by circularly polarized radiation (D1) and probed by linearly polarized radiation (D2).  The two laser beams co-propagate along $x$ and an interferential filter (IF) blocks the pump before the polarimeter (POL) that detects the Faraday rotation of the probe polarization. Three large-size Helmholtz pairs control the static magnetic field (only two of them, $B_{\mathrm{DC}z}$ and $B_{\mathrm{DC}x}$, are shown). Two smaller Helmholtz pairs ($B_{\mathrm{AC}y}$ and $B_{\mathrm{AC}z}$) generate time-dependent components along $y$ and $z$, while a solenoid  ($B_{\mathrm{AC}x}$) is used for the $x$ component. Several quadrupoles (not shown) help counteract static inhomogeneities. }
      \label{fig:setup}
\end{figure}

The setup built to prepare and interrogate the atomic magnetization is described in Ref.\cite{bevilacqua_multi_apb_16}. Briefly (see Fig.\ref{fig:setup}), Cs atoms are optically pumped employing circularly polarized radiation (at mW level) tuned to the D1 line, and probed with a weak ($\mu$W level) linearly polarized radiation detuned by a few GHz from the D2 line. The D1 radiation is brought in resonance for tens of milliseconds and suddenly detuned to start the measurement. A balanced polarimeter is used to detect the Faraday rotation of the probe polarization plane, providing a signal proportional to the sample magnetization component along the probe-beam axis. The two beams co-propagate along the $x$ direction and in magnetometric application a static field is applied along the $z$ direction. The balanced polarimeter uses a transimpedance amplifier with a bandwidth of several kHz, enabling the detection of signals within the audio spectral range.

The experiment is run in an unshielded environment, where the magnetic field and its first-order gradients are controlled by three large-size (180~cm) Helmholtz pairs and five quadrupole sources, all of them driven by numerically controlled direct current (DC) generators that can be set manually or automatically  \cite{biancalana_v2i_rsi_17}. 

\begin{figure}[ht]
   \centering
      \includegraphics [angle=0, width= 0.7 \columnwidth]{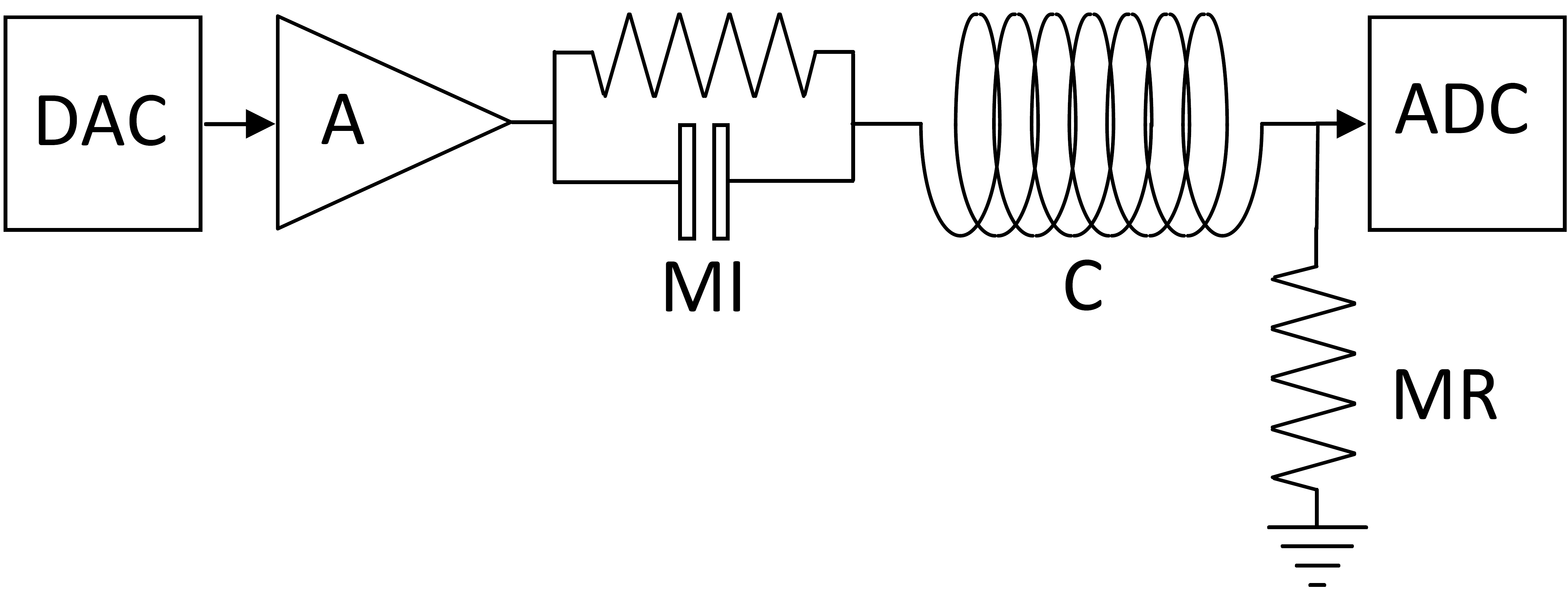}
      \caption{Schematics of the time-dependent field generator and monitor for one Cartesian component of the time-dependent field: DAC: digital-to-analog converter, programmed for arbitrary waveform generation; A: amplifier; MI: matching impedance; C: coil (solenoid or Helmholtz pair); MR: monitor resistance; ADC: analog to digital converter.}
      \label{fig:schema}
\end{figure}

Additional coils enable the application of time-dependent (AC) fields along the three directions, two of which are used in this experiment. The co-propagating arrangement of the pump- and probe-beams permits the use of a solenoid for the $x$ component, while the $y$ and $z$ components are generated by small size (45~cm and 50~cm size, respectively) squared Helmholtz pairs, each with 50 turns per coil. These AC coils are supplied via audio-amplifiers that amplify arbitrary waveforms generated by a digital-to-analog (DAC) converters card (NI~6343), as sketched in Fig.\ref{fig:schema}. Depending on the required frequencies, matching series-impedances can be applied to improve the coupling between each amplifier and the respective coil. The current flowing in each coil is monitored by recording the voltage drop over series resistors, using analog-to-digital converters (ADC) available in the same NI-6343 card.

An \textit{a priori} calibration of all the coils is performed based on size, shape and number of turns using the Biot-Savart law. An \textit{a posteriori} calibration is then performed, under static conditions, by measuring the Larmor frequency in response to assigned DC currents applied to the coils. The Larmor frequency can be evaluated by scanning the frequency of the pump modulation across the magnetic resonance as described in \cite{bevilacqua_multi_apb_16} or estimating the frequency of a free precession signal (FPS) as described, e.g., in \cite{lenci_jpb_12, grujic_epjd_15}.
 
In the case of FPS measurements, the procedure implemented is as follows: the pump diode-laser is tuned to the $F_g=3 \rightarrow F_e=4$ D1 transition for 100~ms, then is abruptly blue-detuned by about 30~GHz by reducing its drive current. As soon as the pump radiation is made off-resonant, the data acquisition starts, lasting another 100~ms. 

\begin{figure}[ht]
   \centering
      \includegraphics [angle=0, width= 75 mm]{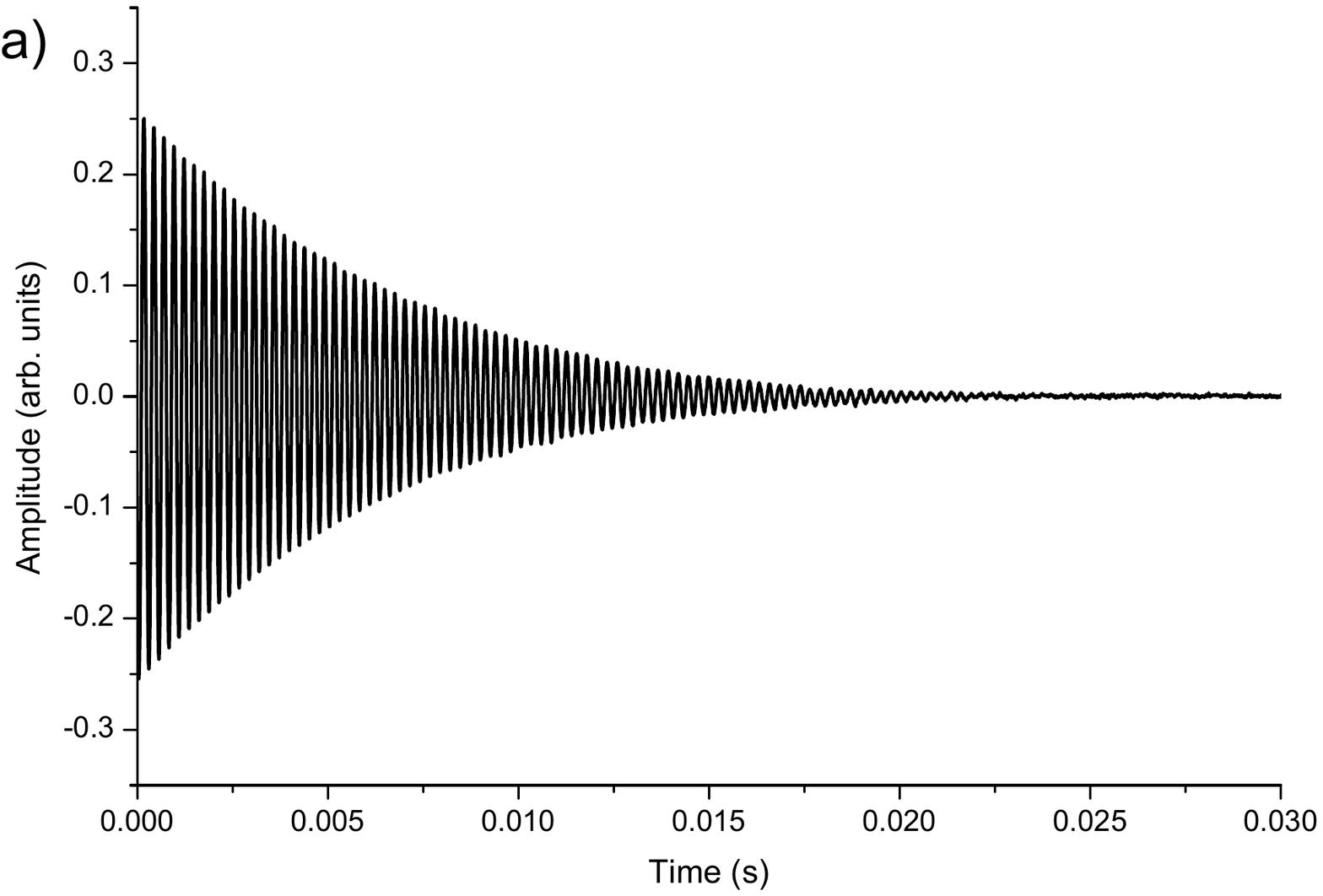}
      \includegraphics [angle=0, width= 75 mm ]{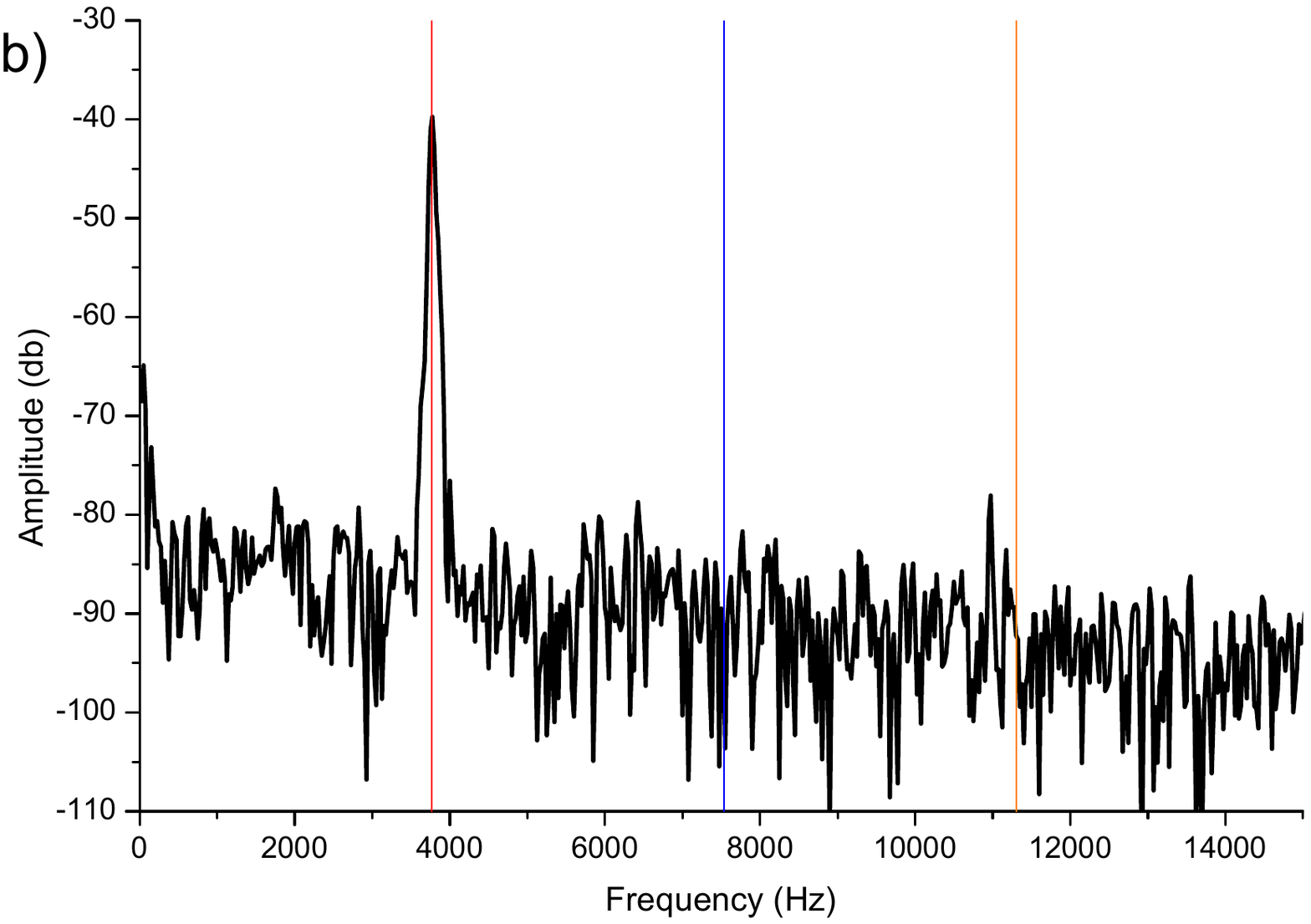}
      \caption{FPS in the time (a) and the frequency (b) domains. The red, blue, and orange lines indicate the precession frequency and its $2^\mathrm{nd}$, an $3^\mathrm{rd}$ harmonics, respectively.} 
      \label{fig:fid}
\end{figure}

An example of such FPS is reported in Fig.\ref{fig:fid}. 
No appreciable harmonic distortion is observed: no harmonic peaks emerge from the noise floor, confirming the linearity of the photo-detection stage. The localization of the fundamental tone peak in Fig.\ref{fig:fid}b permits the evaluation of the static field. In this case, the magnetic field is estimated to be $1.071 \mu$T $\pm 3$nT. A consistent value ($1.070 \mu$T $\pm 3$nT) was obtained from the analysis of the resonance profile obtained in forced conditions, i.e. running the magnetometer in the Bell and Bloom configuration \cite{bevilacqua_multi_apb_16}. The $\pm 3$nT uncertainties are estimated from the standard deviation over large measurement sets, and they are due to the ambient field fluctuations.

The AC field generated by each coil is designed based on the calibration factors determined under static conditions. To this aim, the DAC is programmed to generate a signal with known amplitude and phase at the desired frequency; this signal is amplified and applied to the coil; the voltage drop on the monitor resistor is acquired and analyzed to extract its phase and amplitude; the latter is then converted to current and then to field based on the monitor resistance and the static coil constant; finally the DAC output is scaled and rephased to achieve the field settings (phase and amplitude) set by the operator. Being DAC and ADC of the diverse coils synchronous, this procedure enables the generation of field components with assigned relative phase and amplitudes but relies on the static calibration factors.

Transient signal measurements are also applied to characterize time-dependent, two-dimensional magnetic fields, i.e. for the dynamic calibration procedure at the focus of the present work. In this case, the signal is not a FPS but originates from atomic spins that adiabatically follow the time-dependent applied field. In both cases, the recorded signal is a damped oscillation but in this second case the oscillation frequency is that of the rotating field.

\section{Methodology}
\label{sec:principle}

The proposed methodology requires that the atomic magnetization adiabatically follows the magnetic field to be characterized, i.e., the atomic spins must precess much faster than changes in field orientation. With a time-dependent field that rotates at an angular speed $\omega$, the condition
\begin{equation}
    \gamma B \gg \omega 
    \label{eq:adiab}
\end{equation}
must be fulfilled, where $\gamma$ is the gyromagnetic factor and $B$ is the modulus of the field. 
This requirement may not be compatible with the dressing experiments that inspired this work, for which  $\omega \gg \gamma |B_\mathrm{DC}|$. Nevertheless, the calibration parameters determined with the proposed methodology can subsequently be applied also to operate under dressing conditions.

The calibration factors (valid for an assigned frequency $\omega /2\pi$) are determined by the accurate identification of coil currents (their relative phase and amplitudes) that must be applied to produce a circularly polarized time-dependent field that rotates at $\omega$. Specifically, we
consider a field nominally made of a static component along $z$ and a rotating one on the $xy$ plane. Possible imperfections are considered, which may concern the orientation of the DC field and some degree of ellipticity in the polarization of the time-dependent field. Namely, we consider 

%
\begin{equation}
\vec B = B \hat u_B=B_0 \big( \cos (\omega t)+m_1, (1+\varepsilon) \sin (\omega t + \phi)+m_2, m_3\big),
\label{eq:field}
\end{equation}
with  $m_1, m_2$ accounting for spurious static field components on the polarization plane $xy$, and $\varepsilon , \phi $ for imperfections of the time-dependent field polarization. $B_0$ is the nominal amplitude of the rotating field and $m_3 B_0$ is the nominal amplitude of the static one.
The aim is to obtain a time-dependent field with circular polarization, i.e. $\varepsilon=\phi=0$, on a plane perpendicular to the static term, i.e. $m_1=m_2=0$. 

In the hypothesis of adiabatic following, the macroscopic magnetization $\vec M$ maintains a constant modulus $M$ (apart from the decay due to the relaxation mechanisms) while precessing around $\vec B(t)$. Among the Cartesian components of $\vec M$, the one parallel to $\vec B$  remains approximately constant, while the perpendicular ones oscillate at the high frequency  $\gamma |B|$ \cite{abragam_book_61}. 

Therefore, neglecting the fast oscillating terms, the dynamics of $\vec M$ is determined by the evolution of
\begin{equation*}
    \vec M_{\parallel} = M_{\parallel} \hat u_B 
\end{equation*}
where $M_{\parallel}$ is constant. The $x$ component of $\vec M_{\parallel}$, that is the low-frequency term revealed by our polarimetric detector and used for dynamic coil calibration is
\begin{equation}
    M_{\parallel x}(t) = M_{\parallel} \frac{B_0 \big(m_1+\cos (\omega t)\big)}{B}.
\label{eq:mparallel}
\end{equation}

The proposed procedure consists of a preliminary step aimed at ensuring $m_1=m_2=0$, and a main one to make $\varepsilon=\phi=0$. Both objectives are pursued through Fourier analysis of the Taylor approximation of $M_{\parallel x}(t)$ (eq. \ref{eq:mparallel}) -- relevant details provided in the Appendix.
The preliminary step pertains only to the DC components of the field and could alternatively be performed using traditional methods. The second step, which specifically addresses the AC components, leads to the primary scope of the proposed methodology.

\subsection{Alignment of the static field along $z$ ($m_1=m_2=0$ condition)}
\label{subsec:m1m2}
A straightforward and commonly applied procedure to orient a static field along an axis (let it be $z$, as in our case) is based on minimizing the Larmor frequency by varying the currents that drive the transverse ($x$ and $y$) field components: the minimum is achieved when those transverse components of the field are fully compensated. 

In the presence of a rotating field on the $xy$ plane, harmonic analysis of the detected signal, eq.\ref{eq:app:signal}, can be used to reveal the presence of static components on the $xy$ plane, as well.
Let's analyze the harmonic content of the measured signal described by eq.\ref{eq:mparallel}.
The power ratio between the second-harmonic terms and the fundamental one (eqs.\ref{eq:app:second} and \ref{eq:app:fundam}) is:
\begin{equation}
    A_2=\frac{\langle f_2 \rangle^2}{\langle f_1 \rangle^2}=\frac{m_1 ^2+m_2^2}{8\left(1+m_3^2\right)^2},
    \label{eq:2nd}
\end{equation}
so that minimizing  $A_2$ will lead to the condition $m_1=m_2=0$, i.e. to a static field perpendicular to the polarization plane of the time-dependent components. 
Similar to the case of Larmor frequency minimization, the transverse components appear quadratically in the quantity $A_2$ to be minimized.

If the DC and the AC field components are generated by different coil sets,  misalignments might exist between the $xy$ plane defined by the DC coils and the polarization plane of the time-dependent field. As a consequence, the static field can be made perpendicular either to the $xy$ plane or to the polarization plane. The procedure based on Larmor frequency minimization refers to the former, while the one based on harmonic analysis refers to the latter. The study of coil misalignments goes beyond the scope of this paper, however, we note that the just mentioned feature suggests that using the two methods could help highlight imperfect alignment among AC and DC coils.

As shown in the Appendix, non-zero values of $\phi$ and $\varepsilon$ do not contribute to second harmonic terms, making the described procedure robust to imperfect polarization of the rotating field. Indeed (see eqs. \ref{eq:app:second} and \ref{eq:app:third}), the polarization imperfections parametrized by $\phi$ and $\varepsilon$  produce only odd harmonic terms, as discussed in the next.

\subsection{Refinement of the field circular polarization on the $xy$ plane ($\phi=\varepsilon=0$ condition)}
\label{subsec:epsilonphi}

Let's assume that 
the static field (if any) has been aligned 
along $z$, i.e. let the condition $m_1=m_2=0$ and a generic $m_3$ be achieved. 
This paves the way to proceed with the second step of the proposed calibration method.
Harmonics of the fundamental tone are now only ascribed to polarization imperfections, which, as shown in the Appendix, do contribute to even harmonics of the detected signal.

Making reference to eqs.\ref{eq:app:fundam} and \ref{eq:app:third}, we may derive the power ratio between the third harmonic terms and the fundamental one, finding  
\begin{equation}
      A_3=\frac{\langle f_3 \rangle^2}{\langle f_1 \rangle^2}=\frac{\varepsilon ^2+\phi^2}{32\left(1+m_3^2\right)^2}, 
      \label{eq:a3}
\end{equation}
similarly to the case of eq.\ref{eq:2nd}, minimizing $A_3$ will lead toward the $\varepsilon=\phi=0$ condition, i.e. to a condition of perfectly circular polarization of the time-dependent field.

It is worth examining the possible consequences of imperfect perpendicularity of the static field on the accuracy of this procedure. In fact, besides the analyzed second-harmonic terms, a non-perpendicular static field would cause the emergence of third-harmonic term mimicking polarization issues.  However, as shown in eq.\ref{eq:app:third}, in a first-order Taylor approximation $m_1$ and $m_2$ do not produce such an odd term, and they only contribute when also the second order (eq.\ref{eq:app:thirdh2ord}) is taken into account. In this sense, the procedure aimed to obtain circular polarization for the identification of $\varepsilon$ and $\phi$ is robust concerning the $m_1=m_2=0$ requirement.

\section{Experimental validation}
\label{sec:results}

\subsection {Static field alignment}
\label{subsec:misalignment}
\begin{figure}[ht]
   \centering
      \includegraphics [angle=0, width= 75mm]{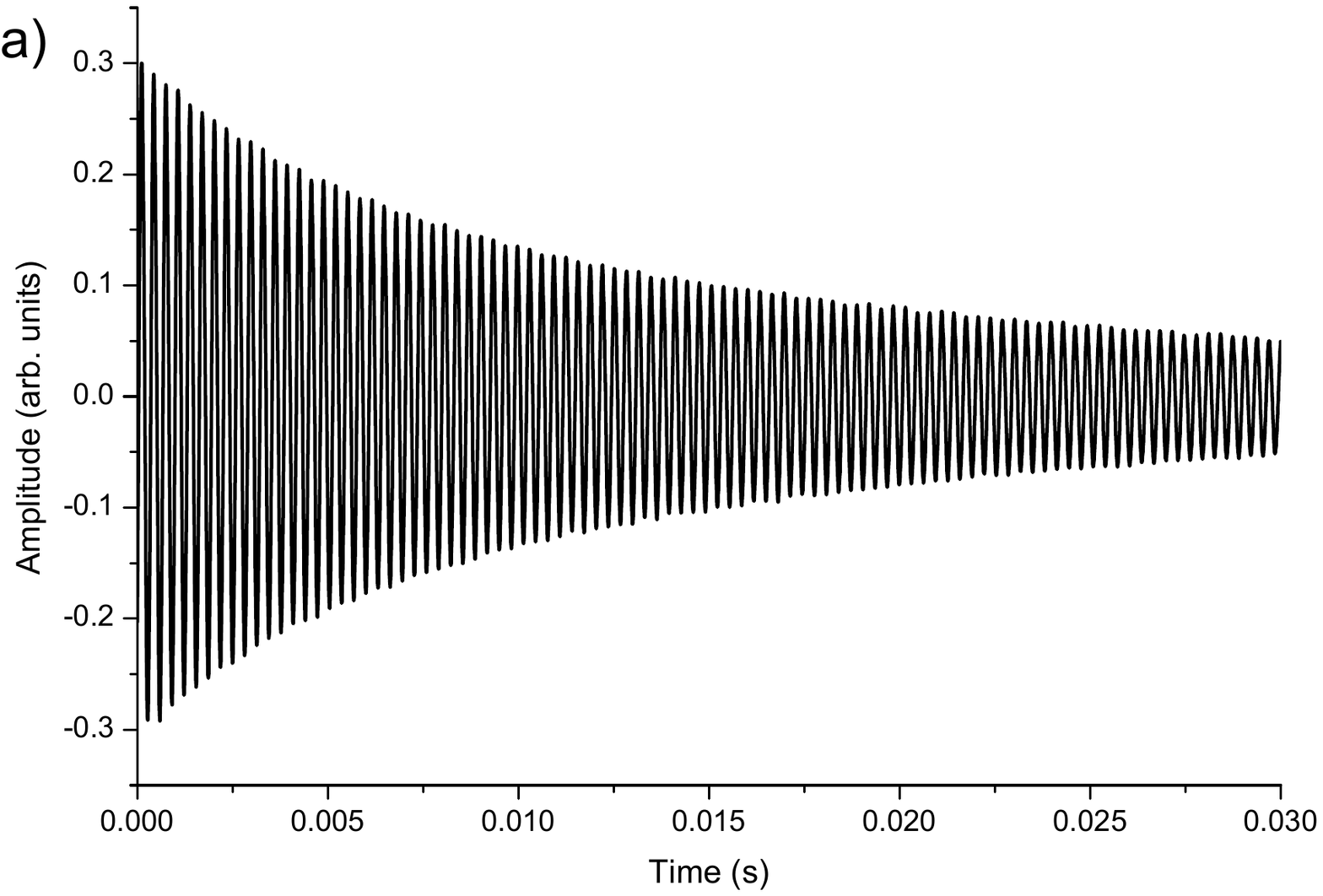}
      \includegraphics [angle=0, width= 75mm]{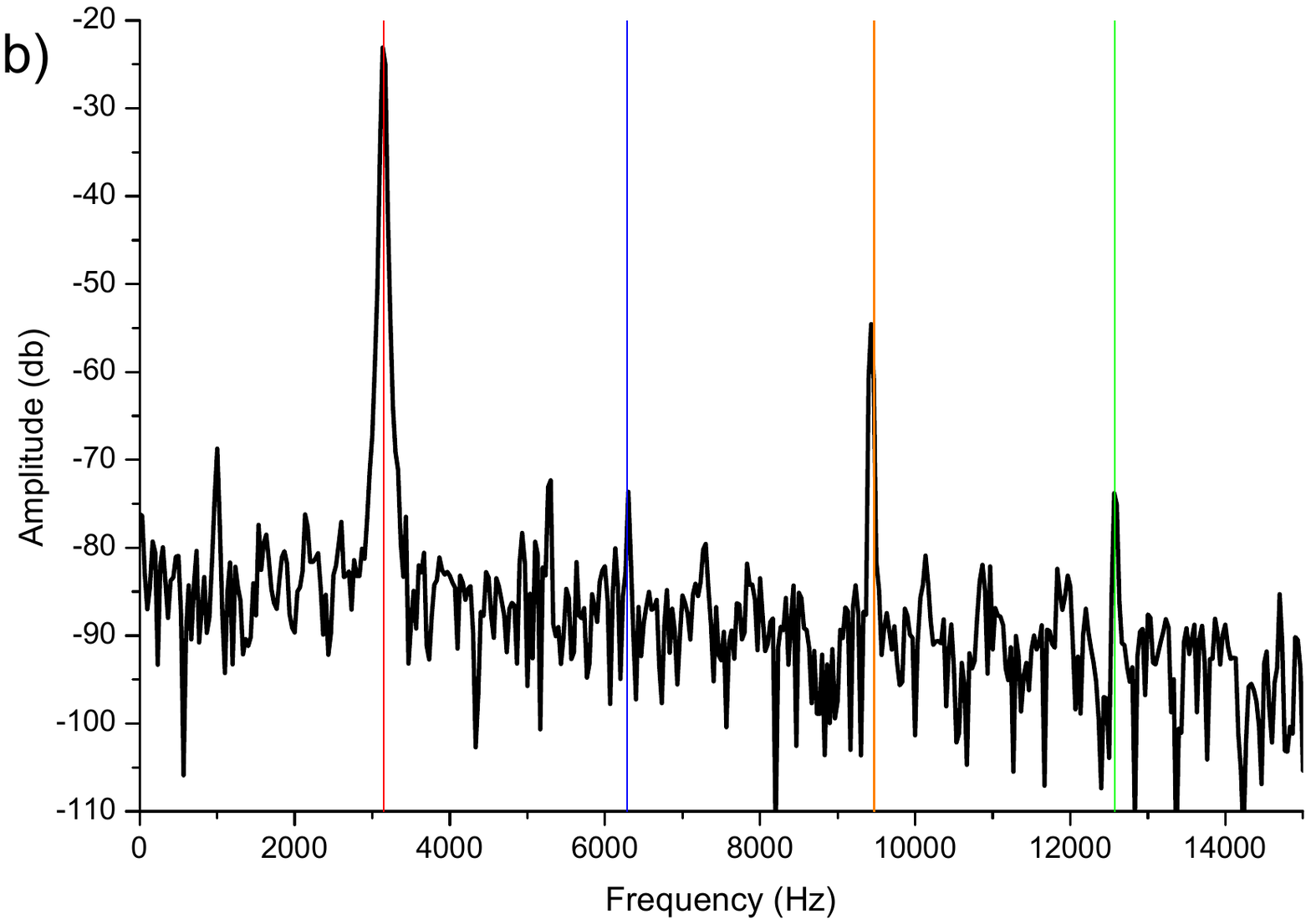}
      \caption{Signal generated from atomic magnetization adiabatic following a 16~$\mu$T  field rotating at 3.144~kHz on the $xy$ plane, while the static field is the same as for Fig.\ref{fig:fid}, in time (a) and frequency (b) domain, respectively. A second harmonic peak is barely recognizable, about 50~dB below the fundamental tone. {The red, blue, orange, and green lines indicate the field rotation frequency and its $2^\mathrm{nd}$, $3^\mathrm{rd}$, and $4^\mathrm{th}$ harmonics, respectively.}
      } 
      \label{fig:no2nd}
\end{figure}

\begin{figure}[ht]
   \centering
      \includegraphics [angle=0, width= 75mm]{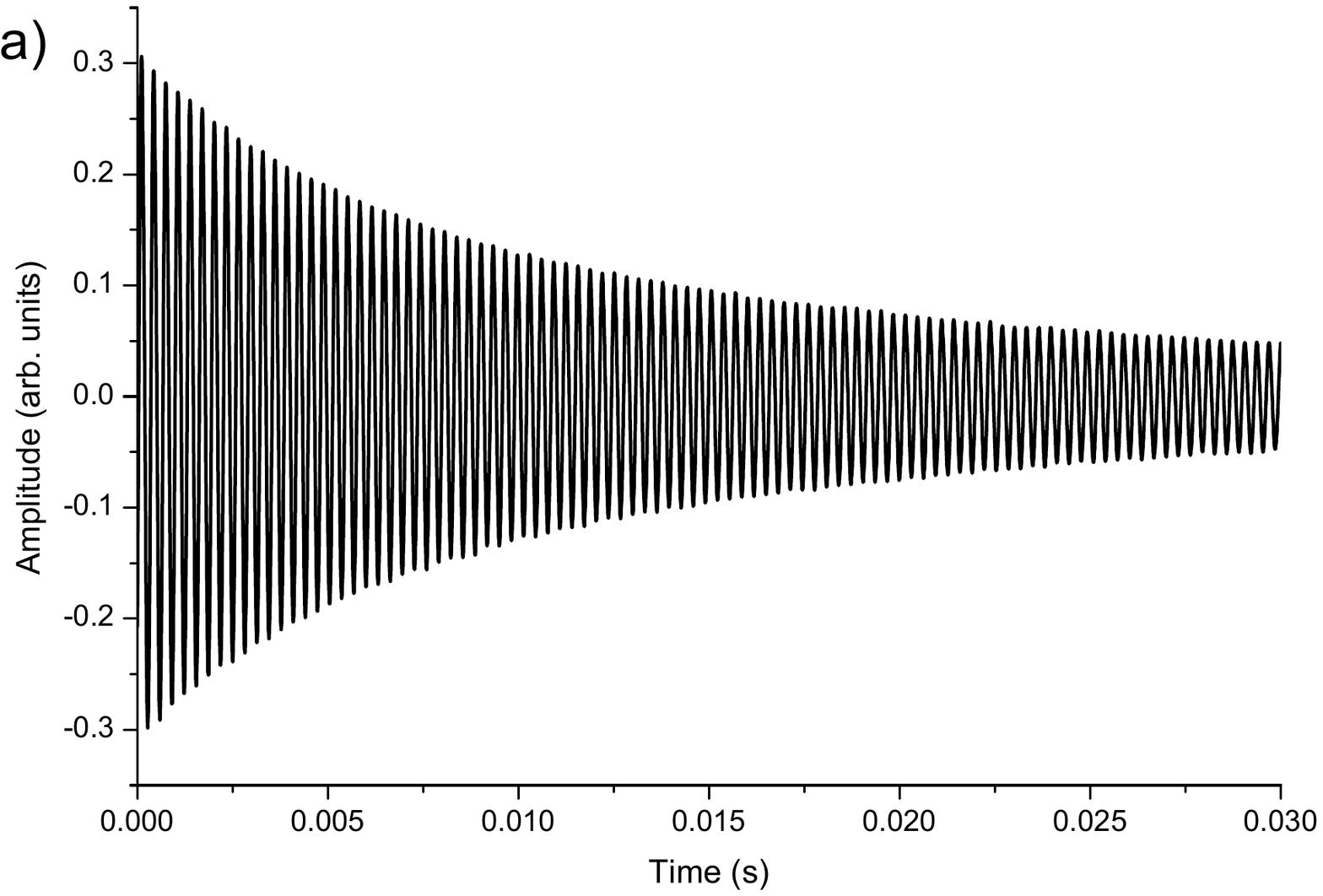}
      \includegraphics [angle=0, width= 75mm]{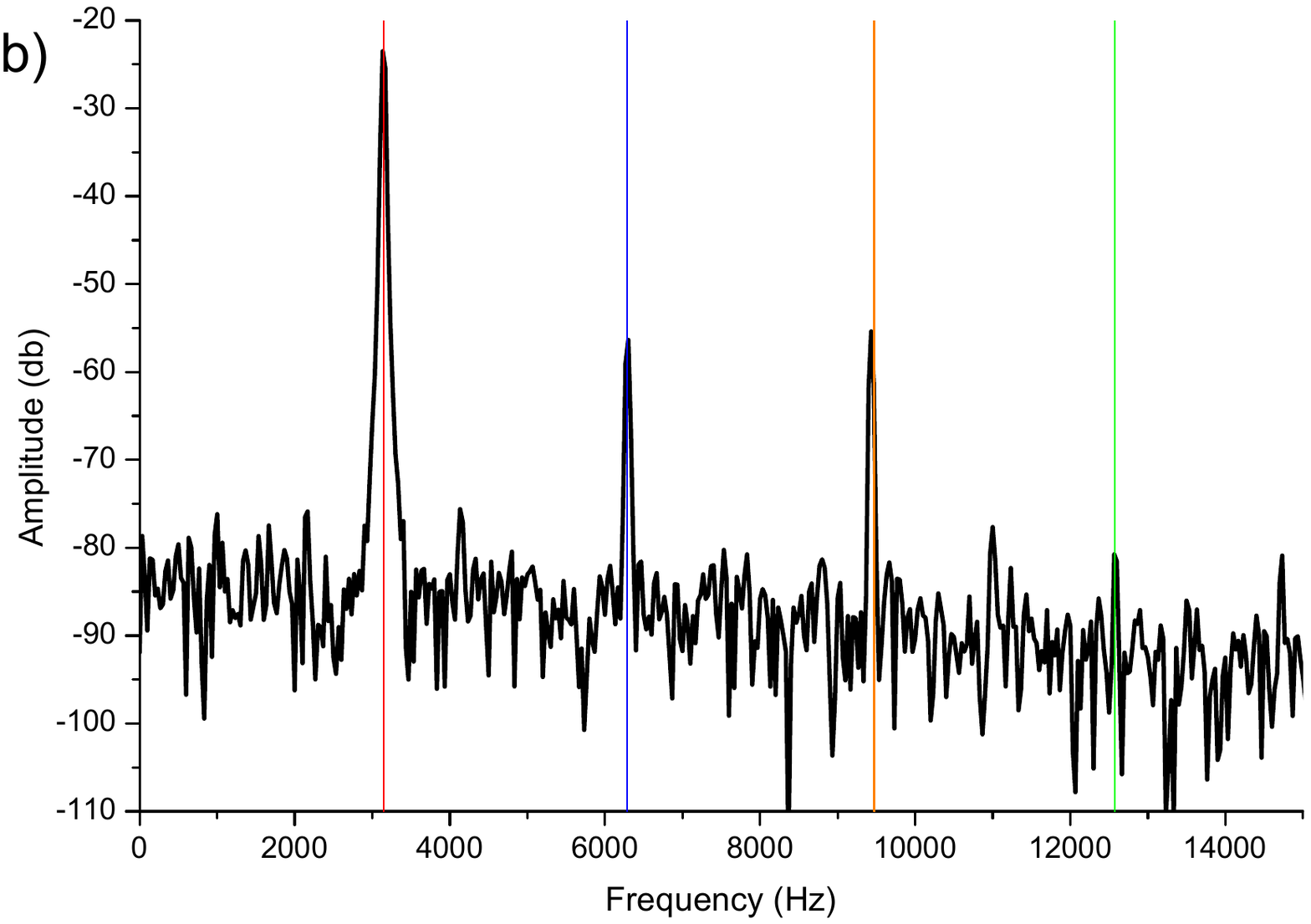}
      \caption{As in Fig.\ref{fig:no2nd}, the magnetization adiabatically follows a 16~$\mu$T,  3.144~kHz field rotating on the $xy$ plane. But a transverse component (830 nT in amplitude) is added to the static field along the $z$ direction. Also in this case the signal is plotted in the time (a) and frequency (b) domains. Compared to Fig.\ref{fig:no2nd}b a second-harmonic peak emerges, in quantitative accordance with the prediction of eq.\ref{eq:2nd}. The red, blue, orange, and green lines indicate the field rotation frequency and its $2^\mathrm{nd}$, $3^\mathrm{rd}$, and $4^\mathrm{th}$ harmonics, respectively. } 
      \label{fig:yes2nd}
\end{figure}

We have measured a FPS after having aligned the field along $z$ by Larmor frequency minimization and in the absence of any rotating field. The static field inferred from the FPS is about 1~$\mu$T. We have then applied a field of 16~$\mu$T rotating at 3144~Hz, thus the nominal condition is $m_1=m_2=0, \, m_3=1/16$, the amplitudes and the relative phase of the oscillating $x, y$ components are set to $\pi/2$ based on the static calibration factors. The recorded signal is reported in Fig.\ref{fig:no2nd}: just a weak second harmonic peak appears, and a third harmonic peak is visible, about 33~dB below the fundamental tone.
We have then applied an additional transverse static component that shifts the free precession frequency to 4950~Hz: this corresponds to have $(m_1, m_2, m_3) \approx (0, 52, 66)\cdot 10^{-3}$, which according to eq.\ref{eq:2nd}  causes a second-harmonic peak about 34~dB below the fundamental tone. This expectation is in perfect accordance with the spectrum shown in Fig.\ref{fig:yes2nd}b. For comparison, the condition of Fig.\ref{fig:no2nd}, with the second-harmonic peak 50~dB weaker than the fundamental tone (hence $A_2 \approx  10^{-5}$) gives $m_1^2+m_2^2 \approx 8\cdot A_2 \approx 8 \cdot 10^{-5}$, this corresponds to a $8 \degree$ misalignment, which would decrease down to $2.6 \degree$ in case of a 60~dB ratio, i.e. if the second-harmonic peak were at the level of the noise floor. It is worth noting that the angular uncertainty is not a reliable measure of the method's accuracy because, for small values of  $m_3$ (as in this case), even minor uncertainties in $m_1, m_2$ result in significant angular misalignments.

Noticeably, the measurements shown in Figs.\ref{fig:no2nd} and \ref{fig:yes2nd} put in evidence also an increase of the decay time compared to the case of free precession around a static field (Fig.\ref{fig:fid}). Such increase is due to mechanisms emerging when the spin dynamics is driven by a rotating fields much stronger than the static one. These mechanisms are currently under investigation and this phenomenon is not further discussed in this work.

\subsection {Effects of polarization ellipticity}
\label{subsec:ellipse}

Rotating fields with variable degrees of ellipticity are applied to compare theoretical and experimental evaluations of $A_3(\varepsilon, \phi)$. In the experiments, the couple $(\varepsilon, \phi)$ is varied around its nominal $(0,0)$ value estimated on the basis of static calibration factors. Discrepancies between static and dynamic calibration factors appear as a translation of the surface $A_3(\varepsilon, \phi)$, in particular as a displacement of its minimum. 

\begin{figure}[ht]
   \centering
      \includegraphics [angle=0, width= 75mm]{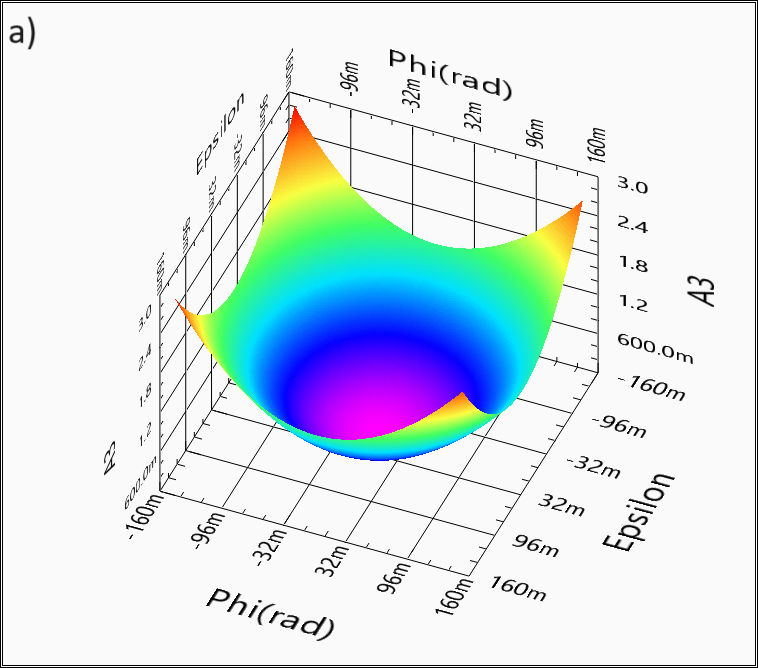} 
      \includegraphics [angle=0, width= 75mm ]{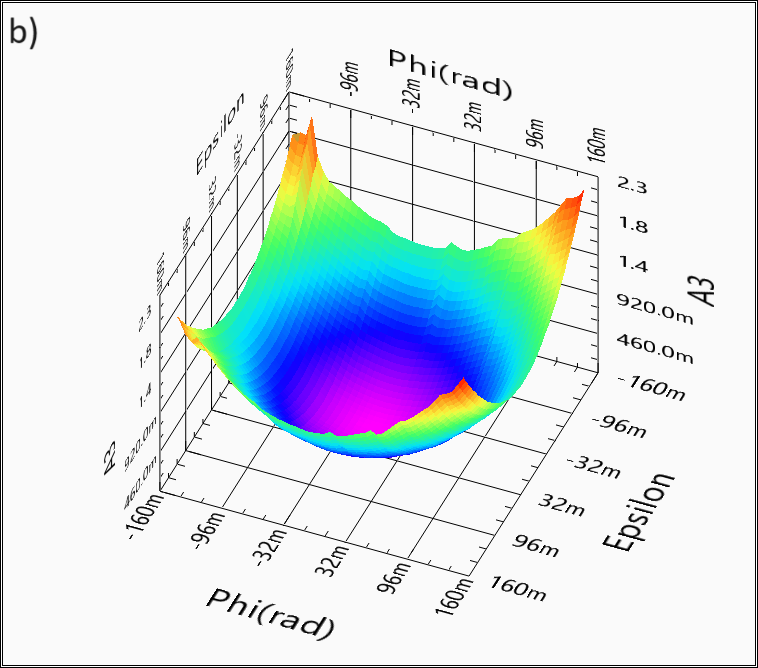}
      \caption{Comparison between calculated (a) and experimental (b) 2D maps for both $\varepsilon$ and $\phi$ spanning a $[-0.15, 0.15]$ range. The experimental one is obtained with $50 \times 50$ measurements of $A_3$, with a low-frequency (1474~Hz) rotating field 16~$\mu$T in amplitude.}
     \label{fig:epsilonphi3d}
\end{figure}

The plots in Fig.\ref{fig:epsilonphi3d} shows a 3D representation of calculated $A_3(\varepsilon, \phi)$ and corresponding experimental results obtained with $m_1=m_2 = 0$. 

The measurements have been performed with 
$\omega=2 \pi \cdot 1474$~rad/s, under the application of a rotating field deliberately distorted with the application of variable couples $(\varepsilon, \phi)$ in the range $[-0.15, +0.15]\times[-0.15 
 \mathrm{~rad}, +0.15  \mathrm{~rad}]$.  The matching is substantial, both in the surface shape and in the absolute values:  the minor deviation (about 13\%) of the latter is due to a low-pass filtering effect by the transimpedance amplifier of the photodetector.

\begin{figure}[ht]
   \centering
      \includegraphics [angle=0, width= 50mm ]{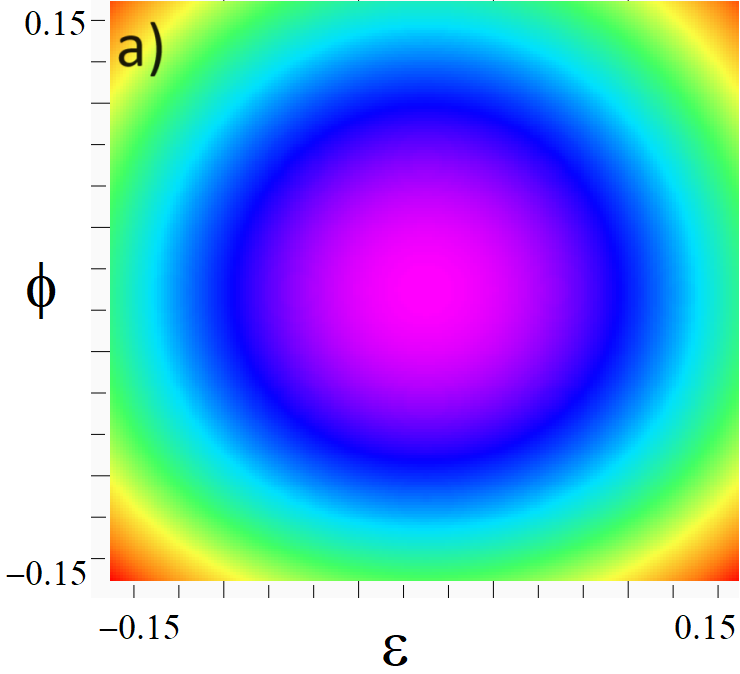} 
      \includegraphics [angle=0, width= 50mm ]{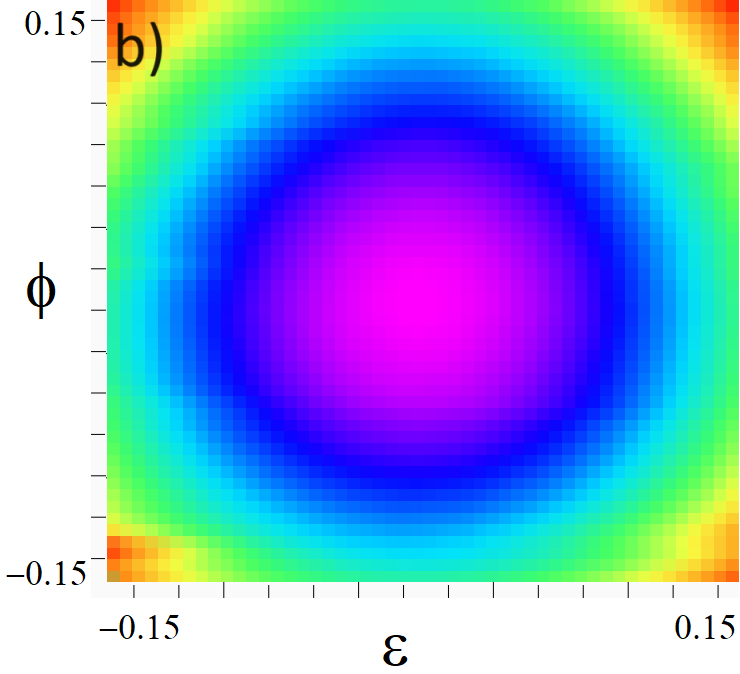}
      \includegraphics [angle=0, width= 50mm ]{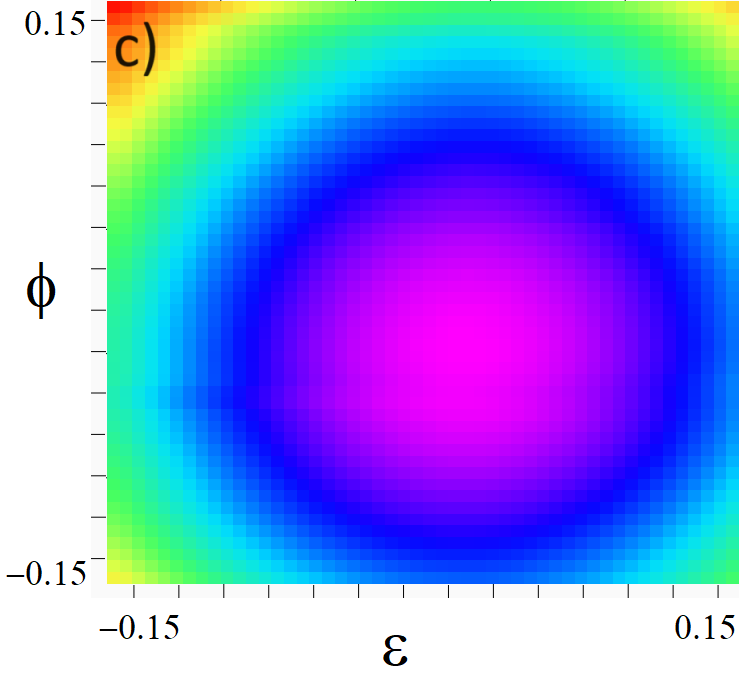}
      \caption{The theoretical and experimental data reported in Fig.\ref{fig:epsilonphi3d} are here represented in the 2D maps a) and b), respectively. The experimental map c) is recorded with a 30~$\mu$T field rotating at a higher frequency (4950~Hz). As expected, the higher frequency causes a larger displacement of the observed minimum from the map's central point, which corresponds to the nominal $(\varepsilon, \phi)=(0, 0)$ based on the static calibration. The displacement of the recorded minimum is ($8\cdot 10^{-4}, 0.35$\degree) in case b) and (0.0195, -1.7\degree) in case c).   }
     \label{fig:epsilonphi2d}
\end{figure}

The same data are then shown as 2D plots in Figs.\ref{fig:epsilonphi2d}a and \ref{fig:epsilonphi2d}b, to facilitate the visualization of the displacement of the minimum. In this case, such displacement is barely detectable: $(\varepsilon_\mathrm{MIN}, \phi_\mathrm{MIN})=(8\cdot 10^{-4}, 0.35 \degree)$.  

Higher frequencies of the rotating field cause larger discrepancies from the static calibration and result in a larger ellipticity and consequently larger displacements of the minimum. The map in Fig.\ref{fig:epsilonphi2d}c is obtained with a 30~$\mu$T field rotating at 4950~Hz. In this case, the minimum of the $A_3$ ratio is located in $(\varepsilon_\mathrm{MIN}, \phi_\mathrm{MIN})=( 0.0195, -1.7\degree )$, and it correspondingly decreases by about 15~dB, down to the level set by the noise floor.

The experimental maps confirm the theoretical prediction and appear with the expected convex surface over wide intervals of the parameters. This suggests that optimization algorithms could be reliably implemented to adjust the relative amplitude and phase of the rotating field to refine the coil calibration and eventually generate a circular polarization at any assigned frequency.

More generally, the method can be applied to identify the phase difference between the drive signals and the magnetic field generated by the corresponding coils, making it possible to produce arbitrary 3D magnetic fields.

\section{Conclusion}
\label{sec:conclusion}
We have introduced a methodology to align static components of a magnetic field and to calibrate its time-dependent components based on \textit{in-situ} measurements performed with a setup for optical-pumping magnetometry.
The procedures are based on the harmonic analysis of polarimetric signals recorded in the presence of an intense, rotating magnetic field and, possibly, of a static one.

We have shown that second harmonic terms can be used to reveal the presence of DC field components lying on the polarization plane of the rotating one. Then we have described a procedure suited to refine the polarization of the latter.
In typical conditions, the first goal can also be pursued with a traditional approach based on Larmor frequency minimization. In contrast, the method developed for polarization refinement constitutes a useful tool to retrieve accurate calibration factors taking into account spurious effects as parasitic capacitance or eddy currents induced in the surroundings, which enable the application of fast-oscillating field components with precisely assigned relative amplitudes and phases. 

We have examined the procedure for the specific case of producing a circularly polarized field on a given plane at a given frequency. An analogous procedure can be implemented on a perpendicular polarization plane and at the diverse frequencies of interest. This would provide a complete set of calibration factors sufficient to generate three-dimensional fields with arbitrary configuration and time dependence. 

\section*{Acknowledgments}
A.Fregosi is pleased to acknowledge the support of EuPRAXIA Advanced Photon Sources project under the contract  EuAPS IR0000030, CUP I93C21000160006. G.Bevilacqua acknowledges the partial support of the MIUR Project PRIN 2020, \textit{Mathematics for Industry 4.0}, Project No. 2020F3NCPX, and of the GNFM-INdAM.

\appendix

\renewcommand{\theequation}{A.\arabic{equation}}
\setcounter{equation}{0}

\section*{Appendix}
\label{sec:appendix}
As discussed in Sec.\ref{sec:principle}, the proposed methodology (developed to point out spurious components of the static field and polarization imperfections of the time-dependent one) is based on the harmonic analysis of the detected signal, which is proportional to the $x$ component of $\hat u_B(t)=\vec B(t)/B(t)$ that is

\begin{equation}
S(t)=\frac
{m_1+\cos (\omega t)} {\sqrt{[m_1+\cos (\omega t)]^2+
[m_2+(1+\varepsilon)\sin (\omega t+\phi)]^2+m_3^2}} 
\label{eq:app:signal}
\end{equation} 

A multivariate Taylor expansion of $S(t)$ in the small variables $m_1, m_2, \varepsilon$ and $\phi$ followed by a Fourier analysis leads to determining the signal components at $\omega$ and its multiples. 

It is convenient to normalize the harmonics terms to the fundamental one, to cancel the effects of spurious amplitude fluctuations of $S(t)$ that may occur, e.g., due to power or tuning fluctuations of the laser sources. Thus we evaluate here in a first-order Taylor approximation the terms at $\omega$, $2 \omega$ and $3 \omega$, which are relevant to the proposed analysis.

The fundamental tone is:
\begin{equation}
    f_1=\frac{\cos(\omega t)}{\left(1+m_3^2\right)^{1/2}}-
    \frac{\varepsilon\cos(\omega t)+\phi\sin(\omega t)}{4\left(1+m_3^2\right)^{3/2}};
    \label{eq:app:fundam}
\end{equation}
the first-order terms oscillating at $2\omega$ only depend  on $m_1$ and $m_2$:
\begin{equation}
    f_2=-\frac{m_1\cos(2\omega t)+m_2\sin(2\omega t)}{2\left(1+m_3^2\right)^{3/2}}, 
    \label{eq:app:second}
\end{equation}
while the third-harmonic ones are expressed by
\begin{equation}
    f_3=\frac{\varepsilon\cos(3\omega t)- \phi\sin(3\omega t)}{4\left(1+m_3^2\right)^{3/2}},
        \label{eq:app:third}
\end{equation}
with no dependence on $m_1, m_2$: the misalignment terms contribute to the third harmonics only at the second order of the Taylor expansion, which reads 
\begin{equation}
    f_{3-2nd}=\frac{\alpha\cos(3\omega t)+ \beta\sin(3\omega t)}{32\left(1+m_3^2\right)^{5/2}},
    \label{eq:app:thirdh2ord}
\end{equation}
with $\alpha=(4m_3^2-5)\varepsilon^2-(8m_3^2+11)\phi^2+12(m_1^2-m_2^2)$ and $\beta=(16m_3^2+10)\varepsilon\phi -24m_1m_2$.

\bibliographystyle{unsrt}

\section*{Bibliogaphy}
\bibliography{3harm}

\end{document}